\documentclass[10pt,twocolumn,showpacs,longbibliography,amsmath,amssymb,osajnl,floatfix,superscriptaddress]{revtex4-1}

\usepackage{graphicx}
\usepackage{dcolumn}
\usepackage{bm}
\usepackage{color}
\usepackage{txfonts}
\usepackage{microtype}

\begin{document}

\author{Feng Wan}	 \affiliation{MOE Key Laboratory for Nonequilibrium Synthesis and Modulation of Condensed Matter, School of Science, Xi'an Jiaotong University, Xi'an 710049, China}	

\author{Yu Wang}  	\affiliation{MOE Key Laboratory for Nonequilibrium Synthesis and Modulation of Condensed Matter, School of Science, Xi'an Jiaotong University, Xi'an 710049, China}

\author{Ren-Tong Guo} 	\affiliation{MOE Key Laboratory for Nonequilibrium Synthesis and Modulation of Condensed Matter, School of Science, Xi'an Jiaotong University, Xi'an 710049, China}

\author{Yue-Yue Chen}\affiliation{Department of Physics, Shanghai Normal University, Shanghai 200234, China}

\author{Rashid Shaisultanov} \affiliation{Max-Planck-Institut f\"{u}r Kernphysik, Saupfercheckweg 1,
	69117 Heidelberg, Germany}

\author{Zhong-Feng Xu} 	\affiliation{MOE Key Laboratory for Nonequilibrium Synthesis and Modulation of Condensed Matter, School of Science, Xi'an Jiaotong University, Xi'an 710049, China}	
	
\author{Karen Z. Hatsagortsyan}\email{k.hatsagortsyan@mpi-hd.mpg.de}
\affiliation{Max-Planck-Institut f\"{u}r Kernphysik, Saupfercheckweg 1,	69117 Heidelberg, Germany}
\author{Christoph H. Keitel}
\affiliation{Max-Planck-Institut f\"{u}r Kernphysik, Saupfercheckweg 1,
	69117 Heidelberg, Germany}
\author{Jian-Xing Li}\email{jianxing@xjtu.edu.cn}
\affiliation{MOE Key Laboratory for Nonequilibrium Synthesis and Modulation of Condensed Matter, School of Science, Xi'an Jiaotong University, Xi'an 710049, China}

\title{High-energy $\gamma$-photon polarization in  nonlinear Breit-Wheeler pair production and $\gamma$-polarimetry}

\date{\today}

\begin{abstract}

The interaction of an unpolarized electron beam with a counterpropagating  ultraintense linearly polarized  laser pulse is investigated in the quantum radiation-dominated regime. We employ a  semiclassical Monte Carlo method to describe  spin-resolved electron  dynamics, photon emissions and polarization, and pair production. Abundant high-energy linearly polarized $\gamma$ photons are generated intermediately during this interaction via nonlinear Compton scattering,  with an average polarization degree of 
more than 50\%, which further interacting with the laser fields produce electron-positron pairs due to nonlinear Breit-Wheeler process. The photon polarization is shown to significantly affect the pair yield by a factor  beyond 10\%.
The considered signature of the photon polarization in the pair's  yield  can be experimentally identified  in a 
prospective two-stage setup. Moreover, the signature can
serve also for the polarimetry of high-energy high-flux $\gamma$ photons with a resolution  
well below 1\% with currently achievable laser facilities.

\end{abstract}

\maketitle

Rapid advancement of strong laser technique enables experimental investigation of  quantum electrodynamics (QED) processes during laser-plasma or laser-electron beam interactions. Nowadays, ultrashort ultrastrong laser pulses can achieve peak intensities of about $10^{22}$ W/cm$^2$, with a duration of about tens of femtoseconds and an energy fluctuation $\sim$ 1\% \cite{Yoon2019,Danson_2019,Gales_2018,ELI,Vulcan,
 Exawatt,CORELS}. In such laser fields  QED processes become nonlinear involving multiphoton processes \cite{Ritus_1985}: a $\gamma$ photon can be generated via nonlinear Compton scattering absorbing millions of laser photons \cite{Goldman_1964, Nikishov_1964, Brown_1964}, or similarly a $\gamma$ photon can create an electron-positron pair in the interaction with a strong laser wave in nonlinear Breit-Wheeler (BW) process \cite{Reiss1962}. These processes have been first experimentally observed in \cite{Bula_1996,Burke_1997,Bamber_1999} and recently  considered in all-optical experimental setups \cite{Sarri2014,Sarri_2015,Yan_2017,Cole_2018,Poder_2018}. Presently, there are many theoretical proposals aiming at $\gamma$ ray and pair production with ultrastrong laser fields of achievable or soon-coming  intensities \cite{Hu_2010, Titov_2012,Hu_2011,Ridgers_2013,Bashmakov_2014, Nakamura_2015, Blackburn_2017, Olugh_2019} and even avalanche-like electromagnetic cascades in  future extreme laser intensities $\gtrsim 10^{24}$~W/cm$^2$ \cite{Marklund2006, Mourou2006,Bell_2008, Piazza2012,Vranic_2016,Tamburini_2017,Gonoskov_2017}.

Recently, it has been realized that the radiation reaction due to $\gamma$ photon emissions in laser fields can be harnessed to substantially polarize electrons \cite{Sorbo_2017,Sorbo_2018,Seipt_2018,Seipt_2019,li2019prl,Song_2019,Li_2019spin} or to create polarized positrons \cite{Chen_2019,Wan_2019plb},
while it was known since long ago that an electron beam cannot be significantly polarized by a monochromatic laser wave \cite{Kotkin2003prstab,Ivanov_2004,Karlovets_2011}.
Polarization properties of electrons, positrons and $\gamma$ photons in ultrastrong laser-electron beam interaction have been investigated comprehensively in \cite{li2019prl,Chen_2019,Ligammaray_2019}.  In particular,
an efficient way of the polarization transfer from electrons to $\gamma$ photons in such interaction has been identified, which will allow to obtain circularly polarized brilliant $\gamma$ rays via nonlinear Compton scattering from longitudinally spin-polarized electrons \cite{Ligammaray_2019}, highly sought  in detecting schemes of vacuum birefringence in ultrastrong laser fields \cite{Nakamiya_2017,Bragin2017}.

In spite of significant efforts in the investigation of  pair production channels in ultrastrong laser-electron beam interaction \cite{Hu_2010, Titov_2012, Ridgers_2013,Bashmakov_2014, Nakamura_2015, Blackburn_2017, Olugh_2019,Bell_2008, Piazza2012, Marklund2006, Mourou2006}, it still remains unclear how the polarization of intermediate particles influences the pair production process in ultrastrong focused laser beams. General theory for the pair production by polarized photons in a monochromatic plane wave  is given in \cite{Kotkin_2005},
obtaining  unwieldy expressions for probabilities which, however, are not directly applicable for processes in tightly focused or multiple laser beams.
 A particular case of the multiphoton BW process with linearly polarized (LP) $\gamma$ photons of a MeV energy and  moderately strong x-ray laser field is considered in  \cite{Krajewska_2012bw}.
The role of the $\gamma$ photon polarization within the BW process in a constant crossed field is considered in \cite{King_2013}, applying a spin-averaged treatment for the photon emission by an electron. While the latter  gives a hint on the  photon polarization effect, it is not straightforwardly extendible to the realistic setups with tightly focused laser beams.

\begin{figure}[t]
  	\setlength{\abovecaptionskip}{-0.0cm}
 	\includegraphics[width=1.0\linewidth]{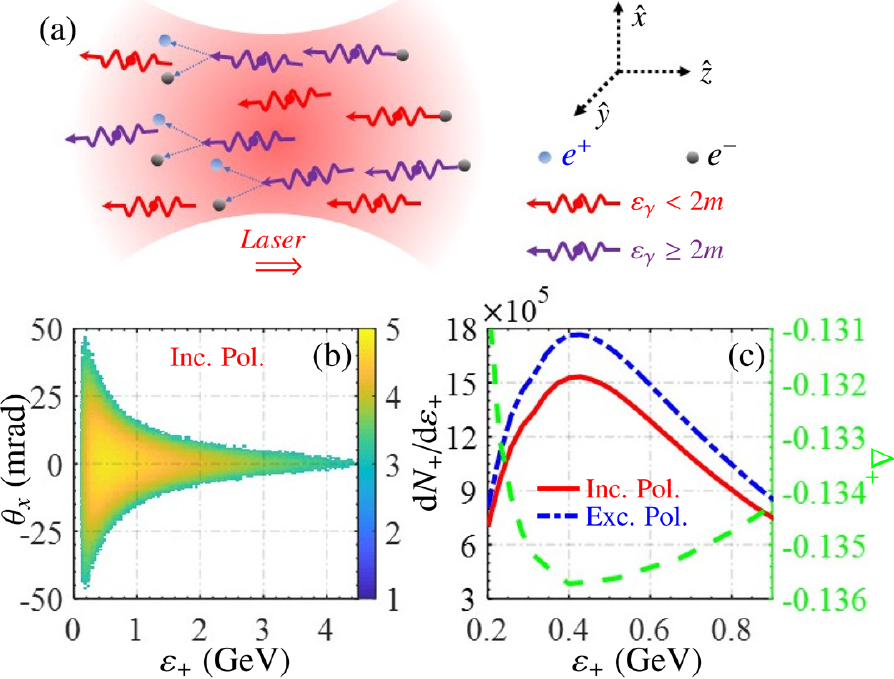}	
 \caption{(a) Scenario of nonlinear BW pair production. A LP laser pulse, polarized in $x$ direction and propagating along $+z$ direction, head-on collides with an electron beam, generating LP $\gamma$ photons,  and further these pairs. ``$e^+$'' and ``$e^-$'' indicate positron and electron, respectively.
  (b) Angle-resolved positron density  log$_{10}$(d$^2N_{+}$/d$\varepsilon_{+}$d$\theta_x$) (GeV$^{-1}\cdot$mrad$^{-1}$) vs the deflection angle $\theta_x=p_{x}/p_z$ and the positron energy $\varepsilon_{+}$, with accounting for the photon polarization.
  (c) Positron density  d$N_{+}$/d$\varepsilon_{+}$ vs $\varepsilon_{+}$
  in the cases of including (red-solid) and excluding (blue-dash-dotted) the photon polarization, respectively.  The green-dashed curve shows the relative deviation $\Delta_{+}$  = [(d$N_{+}$/d$\varepsilon_{+}$)$^{Inc. Pol.}$-(d$N_{+}$/d$\varepsilon_{+}$)$^{Exc. Pol.}$]/(d$N_{+}$/d$\varepsilon_{+}$)$^{Exc. Pol.}$. The laser and electron beam parameters are given in the text (in the paragraph beginning below Eq.~(\ref{pair})).
} \label{fig1}
 \end{figure}

In this Letter, the BW pair production process in a realistic laser-electron beam interaction setup is investigated in the quantum radiation-dominated regime. An  unpolarized ultrarelativistic electron beam is considered to head-on collide with an ultrastrong LP tightly focused laser pulse, which results in radiation of highly LP high-energy $\gamma$ photons via nonlinear Compton scattering.  Further,   generated polarized $\gamma$  photons interact with the laser fields creating electron-positron pairs within the nonlinear BW process; see the interaction scenario in Fig.~\ref{fig1}(a). We apply a fully-polarization-resolved Monte Carlo simulation method developed in \cite{li2019prl,Ligammaray_2019} to describe the spin-resolved electron dynamics, polarized photon emissions, and pair production by the latter.
  We elucidate the substantial role of intermediate polarization of photons on the pair's yield, and 
put forward a two-stage setup for detection of the  photon polarization signature, using laser fields of different linear polarizations and different intensities  in these stages. Moreover, our results suggest  an interesting application in  high-resolution polarimetry of high-energy high-flux LP $\gamma$ rays through the pair  yield.

Note that the high-resolution polarimetry of high-energy $\gamma$ rays is an important problem  in astrophysics and high-energy physics \cite{Lei_1997, McConnell_2006, Eingorn_2018, Ilie_2019}. Current polarimetries for high-energy $\gamma$ photons mainly employ the principles of  Compton scattering and Bethe-Heitler pair production by the Coulomb fields of atoms, with an accuracy of about several percents \cite{Eingorn_2018, Ilie_2019}. The former is not efficient at photon energies larger than $ 100$ MeV because of the kinematic suppression of the Compton rate at large scattering angles, and in the latter the photon flux is restricted severely by the convertor materials \cite{Berlin_1950, Gros_2018}.  Our polarimetry 
concept via nonlinear BW pair production is  specifically designed for high-flux GeV $\gamma$ photons and provides a competitive resolution.

We consider  the quantum radiation-dominated regime, which requires a large  nonlinear QED parameter  $\chi_e\equiv |e|\sqrt{-(F_{\mu\nu}p^{\nu})^2}/m^3\gtrsim 1$ (for electrons and positrons) \cite{Ritus_1985} and
$R\equiv \alpha a_0\chi_e\gtrsim 1$ \cite{Koga2005}.
Significant BW pair production requires the nonlinear QED parameter $\chi_{\gamma}\equiv |e|\sqrt{-(F_{\mu\nu}k_{\gamma}^{\nu})^2}/m^3\gtrsim1$ (for $\gamma$ photon) \cite{Ritus_1985, Baier1998}. Here,
$E_0$ and $\omega_0$ are the laser field amplitude and frequency, respectively,  $p$ and $k_\gamma$ the 4-momenta of  electron (positron) and photon, respectively, $e$ and $m$ the electron charge and mass, respectively, $F_{\mu\nu}$ the field tensor, $\alpha$ the fine structure constant, and $a_0=eE_0/m\omega$ the invariant laser field parameter.
Relativistic units with $c=\hbar=1$ are used throughout.

In our Monte Carlo method, we treat spin-resolved electron dynamics semiclassically, photon emission and pair production quantum mechanically in the local constant field approximation \cite{Ritus_1985,Baier1998,Ilderton2019prd, piazza2019},
valid at $a_0\gg 1$.
At each simulation step, the photon emission is  calculated following the common algorithms~\cite{Ridgers_2014,Elkina2011,Green2015}, and the photon polarization  following the  Monte Carlo algorithm \cite{Ligammaray_2019}.
The photon Stokes parameters  ($\xi_1$, $\xi_2$, $\xi_3$) are defined with respect to the axes $\hat{{\bf e}}_1=\hat{\bf a}-\hat{\bf v}(\hat{\bf v}\hat{\bf a})$ and $\hat{{\bf e}}_2=\hat{\bf v}\times\hat{\bf a}$ \cite{McMaster_1961}, with the photon emission direction $\hat{\bf n}$ along the ultrarelativistic  electron velocity ${\bf v}$,
$\hat{\bf v}={\bf v}/|{\bf v}|$,
and the unit vector $\hat{{\bf a}}={\bf a}/|{\bf a}|$ along the electron acceleration  ${\bf a}$.
After the photon emission the electron spin state is determined by the spin-resolved emission probabilities \cite{li2019prl, CAIN}.
Between photon emissions, the spin precession is governed by the Thomas-Bargmann-Michel-Telegdi
equation \cite{Thomas_1926, Thomas_1927, Bargmann_1959}. The polarized photon conversion to electron-positron pair is described by the probabilities of the pair production. The latter, summing over the pair spins, is derived in the leading order contribution with respect to $1/\gamma_e$ via the QED operator method of Baier-Katkov \cite{Baier_1973}:
\begin{eqnarray}\label{pair}
\frac{{\rm d^2}{W}_{pair}}{{\rm d}\varepsilon_{+}{\rm d}t}&=&\frac{\alpha m^2}{\sqrt{3}\pi\varepsilon_\gamma^2}\left\{{\rm IntK}_{\frac{1}{3}}(\rho)+\left(\frac{\varepsilon_{+}^2+\varepsilon_{-}^2}{\varepsilon_{+}\varepsilon_{-}}-\xi_3\right){\rm K}_{\frac{2}{3}}(\rho)\right\},
\end{eqnarray}
where,  $\varepsilon_{-}$ and $\varepsilon_{+}$ are the energies of created electron and positron, respectively, with the photon energy $\varepsilon_{\gamma}=\varepsilon_{-}+\varepsilon_{+}$, $\rho\equiv 2\varepsilon_{\gamma}^2/(3\chi_{\gamma}\varepsilon_{+}\varepsilon_{-})$,  ${\rm IntK}_{\frac{1}{3}}(\rho)\equiv\int_{\rho}^{\infty} {\rm d}x {\rm K}_{\frac{1}{3}}(x)$, and ${\rm K}_{n}$ is the $n$-order
modified Bessel function of the second kind. In this relativistic setup the emitted $\gamma$ photon is assumed to propagate along the radiating electron momentum,
and the pair along the parent $\gamma$ photon momentum.
Note that averaging over the photon polarization one obtains the known pair production probability $W_{pair}^{Exc. Pol.}$ \cite{Baier1998}, and $W_{pair}= W_{pair}^{Exc. Pol.} -\xi_3W_\xi$.
  When including polarization in Eq.~(\ref{pair}), the Stokes parameters are transformed from the photon emission frame ($\hat{{\bf e}}_1$, $\hat{{\bf e}}_2$, $\hat{\bf n}$) to the pair production frame ($\hat{{\bf e}}_1'$, $\hat{{\bf e}}_2'$, $\hat{\bf n}$), where $\hat{{\bf e}}_1'$ = [${\bf E}- \hat{\bf n}\cdot(\hat{\bf n}\cdot{\bf E})+\hat{\bf n}\times{\bf B}$]/$|{\bf E}- \hat{\bf n}\cdot(\hat{\bf n}\cdot{\bf E})+\hat{\bf n}\times{\bf B}|$ and $\hat{{\bf e}}_2'=\hat{\bf n}\times\hat{{\bf e}}_1'$, with the electric and magnetic field components ${\bf E}$ and  ${\bf B}$; see~\cite{supplemental}.

 The impact of the photon polarization on the BW pair production is quantitatively demonstrated in Figs.~\ref{fig1}(b) and (c).
The employed  laser and electron beam parameters are the following.
A realistic tightly-focused Gaussian LP  laser pulse  \cite{Salamin2002, supplemental} propagates along $+z$ direction (polar angle $\theta_l=0^{\circ}$),   with   peak intensity  $I_0\approx3.45\times10^{21}$ W/cm$^2$ ($a_0=50$),  wavelength $\lambda_0=1$ $\mu$m,  pulse duration $\tau = 15T_0$ with  period $T_0$, and  focal radius $w_0=5$ $\mu$m.
A cylindrical unpolarized electron beam propagates along $-z$ direction (polar angle $\theta_e=180^{\circ}$), with initial kinetic energy $\varepsilon_0=10$ GeV, angular divergence $\Delta\theta = 0.3$ mrad, energy spread $\Delta \varepsilon_0/\varepsilon_0 =0.06$,  beam radius  $w_e= \lambda_0$,  beam length $L_e = 5\lambda_0$,  emittance $\epsilon_e\approx 3\times10^{-4}$  mm$\cdot$mrad,
electron number $N_e=5\times10^6$, and density $n_e\approx 3.18\times10^{17}$ cm$^{-3}$ with
a transversely Gaussian and longitudinally uniform distribution.
The electron beam parameters are typical for laser-plasma acceleration \cite{Leemans_2019}.
 The pair production and radiation reaction are significant at these parameters as
$\chi_e\approx2.47$, Max($\chi_\gamma$) $\approx 2.34$ and $R\approx 1$, while avalanche-like
cascades are suppressed.

Our simulations show that radiated $\gamma$ photons are dominantly LP with an average polarization of $\overline{\xi_3}\approx55.64\%$. The further produced pairs are characterized in Fig.~\ref{fig1}(b). The transverse angular spread of the positrons is about 90 mrad, and the energies are mainly in the region of 0.2 GeV $\lesssim \varepsilon_+\lesssim$ 4.4 GeV. Integrating over the angular distribution, we show the energy distribution of positrons in Fig.~\ref{fig1}(c). When the intermediate photon polarization is accounted for, the  pair (positron) yield  decreases. The relative difference  reaches the maximum of $|\Delta_{+}|\approx13.6\%$  at $\varepsilon_+\approx 0.4$ GeV, and the average relative deviation $\widetilde{\Delta}_+=(N_+^{Inc. Pol.}-N_+^{Exc. Pol.})/N_+^{Exc. Pol.}\approx-13.44\%$.
For the given parameters the positron number is $N_+^{Inc. Pol.}\approx 1.36\times10^{6}\approx N_e \times 27.2\% $, thus, the deviation of about 13.44\% is remarkable and can be measured with current experimental techniques \cite{Alexander_2008, Chen_2015a, abbott2016prl}.

\begin{figure}[t]
\setlength{\abovecaptionskip}{-0.0cm}  	
	\includegraphics[width=1.0\linewidth]{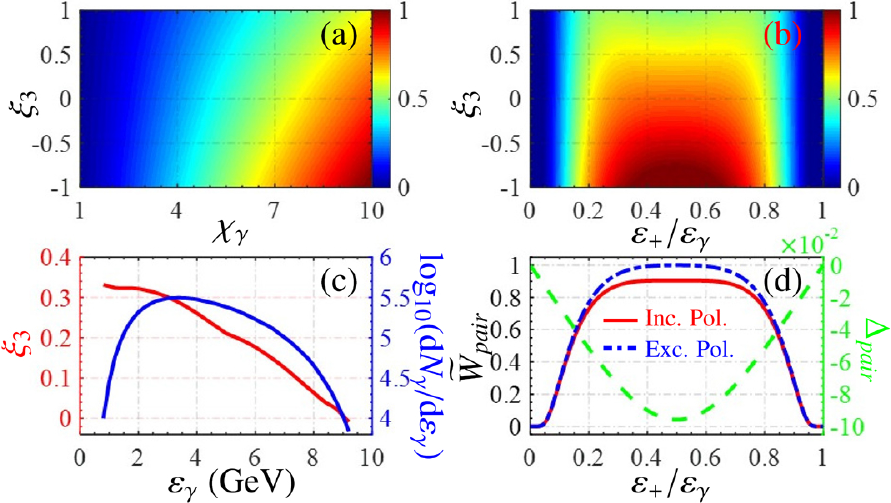}	
		\caption{(a) Normalized pair production probability $\widetilde{W}_{pair}$, integrating over $\varepsilon_{+}$ and scaled by its maximal value at ($\chi_\gamma$, $\xi_3$) = (10, -1) in the demonstrated parametric region, vs $\chi_\gamma$ and  $\xi_3$. (b) $\widetilde{W}_{pair}$ with $\chi_\gamma=2.34$ (corresponding to $\overline{\chi}_\gamma$ of Fig.~\ref{fig1}), scaled by its maximal value at ($\varepsilon_{+}/\varepsilon_\gamma$, $\xi_3$) = (0.5, -1) in the demonstrated parametric region, vs $\varepsilon_{+}/\varepsilon_\gamma$ and $\xi_3$. (c) $\xi_3$ (red) and density log$_{10}$(d$N_{\gamma}$/d$\varepsilon_{\gamma}$) (blue) of emitted $\gamma$ photons, which eventually split to pairs, vs $\varepsilon_\gamma$. (d) $\widetilde{W}_{pair}$  vs $\varepsilon_{+}/\varepsilon_\gamma$ for the cases of including polarization with $\xi_3$ = 25.91\% (red-solid) and excluding polarization (i.e. $\xi_3=0$, blue-dash-dotted), respectively. The green-dashed curve indicates the relative deviation of the pair creation probabilities $\Delta_{pair}$ = ($\widetilde{W}_{pair}^{Inc. Pol.}-\widetilde{W}_{pair}^{Exc. Pol.}$)/$\widetilde{W}_{pair}^{Exc. Pol.}$.  Other laser and electron beam parameters are the same as those in Fig.~\ref{fig1}.}
		\label{fig2}
\end{figure}

\begin{figure}[t]
\setlength{\abovecaptionskip}{-0.0cm}  	
	\includegraphics[width=1.0\linewidth]{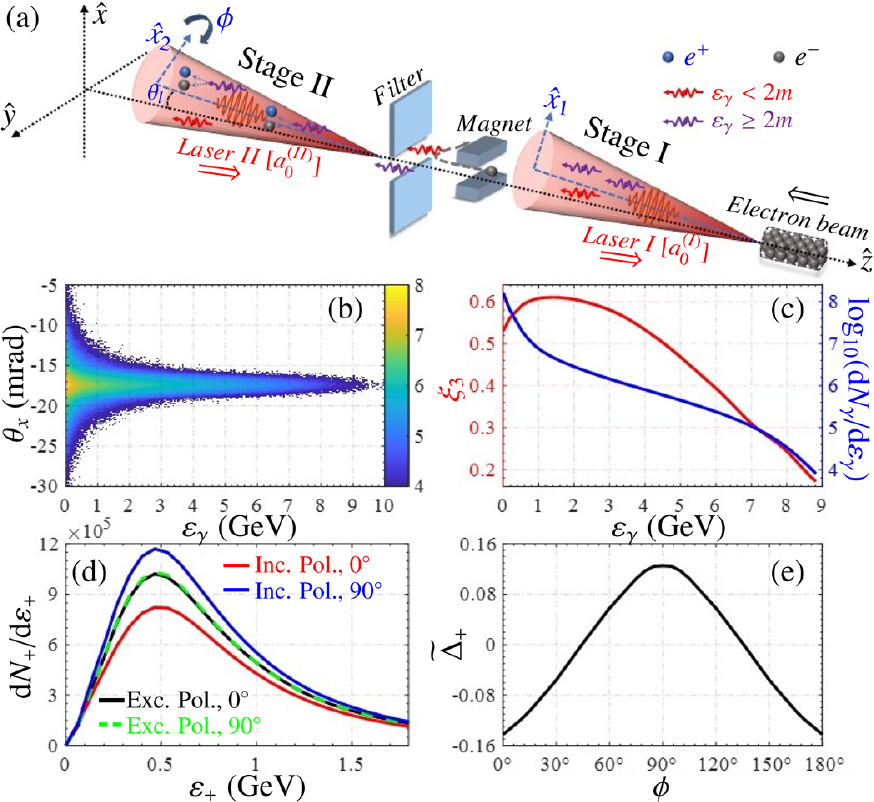}
		\caption{
(a)   Two-stage scenario for detection of the considered effect of the photon polarization. In Stage I and Stage II the laser fields have different LP along $\hat{x}_1$ and $\hat{x}_2$, and different intensities, $a_0^{(I)}=20$ (Max(${\chi}_\gamma$) $\approx0.91$) and $a_0^{(II)}=50$, respectively; $\phi$ is the rotation angle of the polarization $\hat{x}_2$ with respect to  $\hat{x}_2$. 
(b) log$_{10}$(d$^2N_\gamma$/d$\varepsilon_{\gamma}$d$\theta_x$) vs $\theta_x$ and $\varepsilon_\gamma$, generated in Stage I. (c) log$_{10}$(d$N_{\gamma}$/d$\varepsilon_{\gamma}$) (blue), calculated by integrating d$^2N_{\gamma}$/d$\varepsilon_{\gamma}$d$\theta_x$ in (b) over $\theta_x$ 
from -25 mrad to -10 mrad, and the corresponding $\xi_3$ (red) vs $\varepsilon_\gamma$.
(d)
d$N_{+}$/d$\varepsilon_{+}$ vs $\varepsilon_{+}$, produced in Stage II, in the cases of including polarization with $\phi=0^\circ$ (red-solid, $\hat{x}_2\parallel \hat{x}_1$) and $90^\circ$ (blue-solid, $\hat{x}_2\perp \hat{x}_1$) and excluding polarization with  $\phi=0^\circ$ (black-solid) and $90^\circ$ (green-dashed), respectively;  (e) $\widetilde{\Delta}_+$ vs $\phi$.
Other laser and electron beam parameters are the same as those in Fig.~\ref{fig1}.}
		\label{fig3}
\end{figure}

 The physical reason for the intermediate polarization effect is analyzed in Fig.~\ref{fig2}.
According to Eq.~(\ref{pair}), $W_{pair}$ depends on the parameters  $\xi_3$, $\chi_\gamma$ and $\varepsilon_+/\varepsilon_\gamma$. As illustrated in Figs.~\ref{fig2}(a) and (b), $W_{pair}$  continuously decreases (increases) with the increase of
$\xi_3$ ($\chi_\gamma$), and has a symmetric distribution with respect to $\varepsilon_+/\varepsilon_\gamma$.  Intermediate  $\gamma$ photons, which are radiated by the electrons and eventually split to pairs, are LP with an average polarization $\overline{\xi_3}\approx 25.91\%$ (lower than that of all emitted $\gamma$ photons), as demonstrated in Fig.~\ref{fig2}(c). And, the corresponding pair production probability is smaller than that  excluding polarization, in particular, in the region of $0.2 \lesssim \varepsilon_+/\varepsilon_\gamma\lesssim 0.8$ in Fig.~
\ref{fig2}(d). Consequently, the pair yield of consistently including the photon polarization is much smaller than that   with  averaging over the polarization, as shown in Fig.~\ref{fig1}(c).

This photon polarization effect is robust with respect to the laser and electron beam parameters. 
As the laser field parameter $a_0$ varies from $40$ to $60$, the laser
pulse duration from $12$ to $18$ cycles, and the initial kinetic energy of the electron beam $\varepsilon_0$  from 8 GeV to 10 GeV,  the pair production parameter $\widetilde{\Delta}_+$ changes less than 10\% \cite{supplemental}. 
It keeps almost identical for the cases of larger angular divergence of $\Delta\theta=1$ mrad,
larger energy spread $\Delta\varepsilon_0/\varepsilon_0=0.1$ and different colliding angle $\theta_e=175^\circ$ \cite{supplemental}. In the case of employing a circularly polarized  laser pulse, the average polarization  of emitted $\gamma$ photons by the unpolarized electron beam is rather small, and consequently, the considered effect can not be identified \cite{supplemental}.

For experimental verification of the considered effect of the photon polarization, we 
introduce an all-optical two-stage method. In both stages LP laser pulses are used, however, with different polarization directions. In Stage I, a relatively low laser intensity $a_0^{(I)}=20$ (Max(${\chi}_\gamma$) $\approx 0.91$) is used for $\gamma$ photon production via nonlinear Compton scattering (see Fig.~\ref{fig3}(b)) and suppressing the pair creation, while in Stage II a higher laser intensity $a_0^{(II)}=50$ for pair production via the nonlinear BW process. When the laser polarization direction in Stage II is parallel to that in Stage I ($\phi=0^\circ$, $\hat{x}_2\parallel \hat{x}_1$), the pair yield of including polarization  is much smaller than that  excluding polarization, $N_+^{Inc. Pol.}<N_+^{Exc. Pol.}$, with $ \widetilde{\Delta}_+\approx-14.23\%$; see Fig.~\ref{fig3}(d), because ${\xi_3}$ in this frame is positive with $\overline{\xi_3}\approx55.54\%$; see Fig.~\ref{fig3}(c). When the polarization direction in Stage II is rotated by $\phi=90^\circ$, ${\xi_3}$ of $\gamma$ photons in the rotated frame becomes negative, $\overline{\xi_3}\approx-55.54\%$. Consequently, we have $N_+^{Inc. Pol.}>N_+^{Exc. Pol.}$, with $\widetilde{\Delta}_+\approx 12.52\%$; see Fig.~\ref{fig3}(d).
It is clear that  in the case of neglecting the photon polarization, the rotation of the laser polarization in Stage~II would not affect the pair yield. We can explain also why the absolute value $|\widetilde{\Delta}_+|$ in the case of $\phi=0^\circ$  is slightly larger than that of $\phi=90^\circ$. The reason is that the pair production probability is $W_{pair}= W_{pair}^{Exc. Pol.}-\xi_3 W_\xi$ in a single formation length, but within $n$ formation lengths it is $W_n=1-(1-W_{pair})^n$ = $1-[1-(W_{pair}^{Exc. Pol.}-\xi_3 W_\xi)]^n$, which is asymmetric with respect to $\xi_3$. Thus, the dependence of $\widetilde{\Delta}_+$ on the  rotation angle $\phi$  demonstrated in Fig.~\ref{fig3}(e) can be a measurable experimental signature of the considered photon polarization effect.

\begin{figure}[t]
\setlength{\abovecaptionskip}{-0.0cm}  	
	\includegraphics[width=0.95\linewidth]{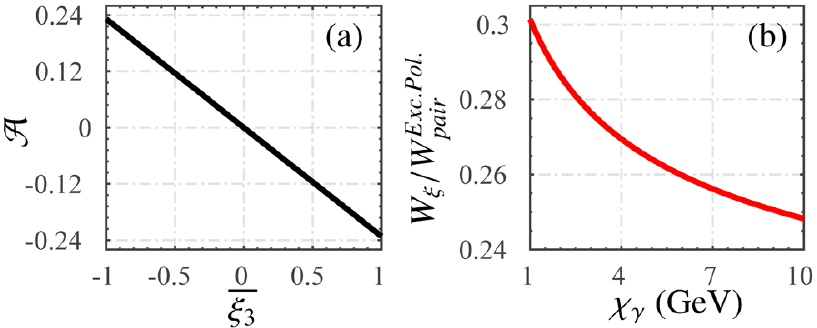}
		\caption{Polarimetry for high-flux high-energy $\gamma$ photons.   (a) The asymmetry parameter ${\cal A}$, defined in the text, vs $\overline{\xi_3}$.    (b) $W_\xi/W_{pair}^{Exc. Pol.}$ vs $\chi_\gamma$ by summing over $\varepsilon_+/\varepsilon_\gamma$.
		The average energy of $\gamma$ photons $\overline{\varepsilon_\gamma}=4.32$ GeV (corresponding to $\overline{\varepsilon_\gamma}$ in Fig.~\ref{fig1}), with an angular divergence $\Delta\theta_\gamma=$ 0.3 mrad and an energy spread $\Delta\varepsilon_\gamma/\varepsilon_\gamma= 6\%$. The scattering laser  parameters are the same as those in Fig.~\ref{fig1}.}
		\label{fig4}
\end{figure}

Finally, we 
suggest a new polarimetry method for high-flux multi-GeV $\gamma$ photons by employing the polarization properties within  the nonlinear BW pair production process.
A $\gamma$ photon beam head-on collides with an ultrastrong  LP laser pulse, and the interaction scenario is similar to Stage II in Fig.~\ref{fig3}(a). The procedure of determining the LP Stokes parameters $\overline{\xi_1}$ and $\overline{\xi_3}$ of the given photon beam is the following.  For the  $\overline{\xi_3}$ determination,  the laser polarization  first is fixed along $x$ direction, and  the positron (pair) yield $N_+|_{\phi=0^\circ}$ is measured.  Then, the laser polarization  is rotated by $90^\circ$, and  again $N_+|_{\phi=90^\circ}$ is measured, which is different from  $N_+|_{\phi=0^\circ}$ since $\overline{\xi_3}$ changes with the rotation of the laser polarization; see similar interpretation in Fig.~\ref{fig3}. Thus,  $\overline{\xi_3}$ can be deduced by an asymmetry parameter
\begin{equation}
\label{asymmetry}
 {\cal A}  = \frac{N_{+}|_{\phi=0^\circ}-N_{+}|_{\phi=90^\circ} }{N_{+}|_{\phi=0^\circ}+N_{+}|_{\phi=90^\circ}},
\end{equation}
and, the relation of ${\cal A}$ to $\overline{\xi}_3$ is shown in Fig.~\ref{fig4}(a).
In the same way the Stokes parameter $\overline{\xi_1}$ can be determined via another asymmetry parameter ${\cal A}'$, first fixing the laser polarization along the axis of $\phi=45^\circ$ and then rotating by $90^\circ$ ($\phi=135^\circ$).

The resolution of the polarization measurement can be estimated
via the statistical uncertainty $\delta {\cal A}/\Delta{\cal A}$ = $1/(\Delta{\cal A}\sqrt{N_+})$  \cite{Placidi_1989}, where the total number of pairs $N_+={\cal R}_{pair} N_\gamma$ is determined by the pair production rate ${\cal R}_{pair}\approx 37.98\%$ and $\Delta{\cal A}$  = Max(${\cal A}$) - Min(${\cal A}$) $\approx0.4634$, calculated with the given parameters.  For instance, in the case of the laser-driven polarized $\gamma$ rays \cite{Ligammaray_2019}, we have $N_\gamma\sim 10^6$, and the resolution is about 0.35\%. As the photon flux increases, the resolution increases accordingly. The resolution improves as well with the increase of the pair yield, which takes place when increasing $\chi_\gamma$ (see analysis in Fig.~\ref{fig2}), and with the increase of the asymmetry parameter $\Delta{\cal A}\sim W_\xi/W_{pair}^{Exc. Pol.}$. The latter, however, decreases with larger $\chi_\gamma$ (see  Fig.~\ref{fig4}(b)). Due to opposite behaviours of $N_+$ and $\Delta{\cal A}$ with the variation of $\chi_\gamma  \propto a_0\varepsilon_\gamma$, the resolution is quite stable with respect to the changes of the laser intensity and the $\gamma$ photon energy.
Moreover, the resolution does not vary much and remains well below 1\% with a shorter or longer laser pulse, a larger energy spread $\Delta\varepsilon_\gamma/\varepsilon_\gamma=0.1$,  a larger angular divergence $\Delta\theta_\gamma$ = 1 mrad, and  a different colliding angle $\theta_\gamma=175^\circ$ \cite{supplemental}.

In conclusion, the impact of intermediate photon polarization on
nonlinear BW pair production during LP laser-electron beam interaction is investigated in the quantum radiation-dominated regime. 
The photon polarization is shown to significantly affect the pair yield by a factor of above 13\%. We 
put forward an all-optical method to experimentally determine the considered signature of the photon polarization. Moreover, we provide a new polarimetry method for high-flux high-energy $\gamma$-rays (in the GeV range) , which provides
 competitive resolution with currently feasible laser facilities,  and is likely to be useful in astrophysics and high-energy physics.\\

This work is supported by the National Natural Science Foundation of China (Grants Nos. 11874295, 11875219 and 11905169), and the National Key R\&D Program of China (Grant No. 2018YFA0404801).

\bibliography{QEDspin}

\end{document}